\documentclass[secnum,leqno]{rims-bessatsu}
\usepackage{amssymb,amsmath}

\newtheorem{thm}{Theorem}[section]

\newtheorem{prp}[thm]{Proposition}

\theoremstyle{definition}

\theoremstyle{remark}
\newtheorem*{rem}{Remark}
\newcommand{\beq}{\begin{equation}}
\newcommand{\eeq}{\end{equation}}
\newcommand{\beqnn}{\begin{equation*}}
\newcommand{\eeqnn}{\end{equation*}}
\newcommand{\rd}{\partial}

\newcommand{\diag}{\operatorname{diag}}

\newcommand{\CC}{\mathbb{C}}
\newcommand{\PP}{\mathbb{P}}
\newcommand{\RR}{\mathbb{R}}
\newcommand{\ZZ}{\mathbb{Z}}
\newcommand{\calL}{\mathcal{L}}
\newcommand{\calP}{\mathcal{P}}
\newcommand{\calT}{\mathcal{T}}
\newcommand{\bsc}{\boldsymbol{c}}
\newcommand{\bst}{\boldsymbol{t}}
\newcommand{\bsx}{\boldsymbol{x}}
\newcommand{\bszero}{\boldsymbol{0}}

\title{Integrable structures of specialized hypergeometric tau functions}

\author{\textsc{Kanehisa Takasaki}
\footnote{Department of Mathematics, Kindai University, 
3-4-1 Kowakae, Higashi-Osaka, Osaka 577-8502, Japan.\newline
e-mail \texttt{takasaki@math.kindai.ac.jp} }}


\support{Supported by JSPS Kakenhi Grant Number JP18K03350.}  


\begin{document}

\maketitle

\begin{abstract}
Okounkov's generating function of the double Hurwitz numbers 
of the Riemann sphere is a hypergeometric tau function of 
the 2D Toda hierarchy in the sense of Orlov and Scherbin.  
This tau function turns into a tau function of the lattice KP 
hierarchy by specializing one of the two sets of time variables 
to constants.  When these constants are particular values, 
the specialized tau functions become solutions of various 
reductions of the lattice KP hierarchy, such as the lattice 
Gelfand-Dickey hierarchy, the Bogoyavlensky-Itoh-Narita 
lattice and the Ablowitz-Ladik hierarchy.  These reductions 
contain previously unknown integrable hierarchies as well. 

\end{abstract}

\section{Introduction}

The notion of hypergeometric tau functions was coined 
by Orlov and Scherbin in an attempt to capture 
multi-variate hypergeometric series in the language 
of the KP and 2D Toda hierarchies \cite{OS01a,OS01b}. 
Other sources of this type of special tau functions 
were the $c = 1$ string theory \cite{DMP93,NTT95,Takasaki96} 
and the Hurwitz numbers of the Riemann sphere \cite{Okounkov00}.  
Since the work of Orlov and Scherbin, 
many applications of hypergeometric tau functions 
have been found in random matrix theory, 
topological string theory, combinatorics and 
enumerative geometry (see the references cited 
in our recent review \cite{Takasaki18a}). 

In this paper, we search for integrable structures 
that underlie a family of hypergeometric tau functions 
of the lattice KP hierarchy (aka the modified KP hierarchy) 
\cite{UT84,Dickey-book}.  These tau functions are obtained 
from a special hypergeometric tau function $\calT(s,\bst,\bar{\bst})$ 
of the 2D Toda hierarchy by freezing the second set 
of time variables $\bar{\bst} = (\bar{t}_k)_{k=1}^\infty$ 
(the so called ``negative'' time variables) 
to a set of constants $\bsc = (c_k)_{k=1}^\infty$. 
$\calT(s,\bst,\bar{\bst})$ is the tau function constructed 
by Okounkov as a generating function of the double Hurwitz numbers 
of the Riemann sphere \cite{Okounkov00}. 
As far as $c_k$'s are generic values, $\calT(s,\bst,\bsc)$
\footnote{To me more precise, we consider $\calT(s,\bst,-\bsc)$ 
rather than $\calT(s,\bst,\bsc)$.}  
is merely a special tau function of the lattice KP hierarchy.  
If, however, $c_k$'s are particular values, $\calT(s,\bst,\bsc)$ 
can be a solution of a reduction (i.e., a subsystem) 
of the lattice KP hierarchy.  

We examine four cases of such special choices of $\bsc$. 
They are known to have many implications for 
the Schur polynomials (or the Schur functions) \cite{Mac-book}.  
The special values of the Schur polynomials for these $\bsc$'s 
play important roles in the work of Orlov and Scherbin 
as well \cite{OS01a,OS01b,Orlov05}.  
The specialized hypergeometric tau functions are related 
to our previous work on topological string theory 
\cite{TN15,TN16} and our recent work on the Hurwitz numbers 
and the Hodge integrals \cite{Takasaki18b,NT19,Takasaki19}.  
We found therein that integrable hierarchies 
of the Ablowitz-Ladik, Volterra and Gelfand-Dickey types 
emerge as underlying integrable structures.  
Remarkably, the specialized hypergeometric tau functions 
indicate the existence of further variations 
of these well known integrable hierarchies.  
We are thus led to a wide spectrum of 
integrable structures hidden behind 
Okounkov's hypergeometric tau function.  

All these integrable structures are realized as reductions 
of the lattice KP hierarchy.  Such a reduction 
is characterized by a particular shape of the Lax operator. 
To find the shape of the Lax operator of a specialized 
hypergeometric tau function, we use the method 
developed in our study on integrable structures 
of melting crystal models and topological string theory 
\cite{Takasaki13a,Takasaki13b,Takasaki14}. 
This method is based on a factorization problem \cite{Takasaki84}.  
The factorization problem enables us to find the initial value 
of the dressing operator at $\bst = \bszero$. 
The initial value of the Lax operator can be computed 
from the initial value of the dressing operator.  
We can thereby identify the relevant reduction. 

This paper is organized as follows.  
Section 2 and 3 are a brief review of the lattice KP hierarchy 
and the notion of hypergeometric tau functions. 
Section 4 is devoted to the factorization problem 
and its implications.  The initial values of the dressing 
and Lax operators of the specialized hypergeometric 
tau functions are computed therein.  The four types 
of specialization of $c_k$'s are also examined in detail.  
The integrable structures hidden behind these initial values 
are formulated in a general form in Sections 5, 6, 7 and 8. 
Sections 5 and 6 deal with integrable hierarchies 
of the Gelfand-Dickey and Volterra types.   
Sections 7 and 8 are focussed on generalizations 
and variations of the Albowitz-Ladik hierarchy.

\section{Lattice KP hierarchy}

Let $\bst = (t_k)_{k=1}^\infty$ be a set of time variables, 
$s$ a spacial variable, and $\Lambda$ the shift operator 
\[
  \Lambda = e^{\rd_s}, \quad \rd_s = \rd/\rd s. 
\]
The Lattice KP hierarchy may be thought of as a subsystem 
of the 2D Toda hierarchy with the flows with respect 
to $\bar{\bst} = (\bar{t}_k)_{k=1}^\infty$ being suppressed.  
In the conventional setup of these integrable hierarchies 
\cite{UT84,Dickey-book}, $s$ is a lattice coordinate, 
hence a discrete variable.  In the following, however, 
we treat $s$ as a continuous variable.  

The Lax formalism of the lattice KP hierarchy is given 
by the Lax equations
\[
  \frac{\rd L}{\rd t_k} = [B_k,L],\quad 
  k = 1,2,\ldots,
\]
for the difference (or, so to speak, pseudo-difference) Lax operator 
\[
  L = \Lambda + \sum_{n=1}^\infty u_n\Lambda^{1-n}, \quad 
  u_n = u_n(s,\bst). 
\]
The generators $B_k$ of the flows are defined 
by the Lax operator as 
\[
  B_k = (L^k)_{\ge 0}  
\]
where $(\quad)_{\ge 0}$ denotes the projection 
\[
  \left(\sum_{n\in\ZZ}a_n\Lambda^n\right)_{\ge 0}
  = \sum_{n\ge 0}a_n\Lambda^n 
\]
to the non-negative power part of a difference operator. 
One can rewrite the Lax equations as 
\[
  \frac{\rd L}{\rd t_k} = [L, B_k^{-}],\quad 
  B_k^{-} = (L^k)_{<0}. 
\]
$(\quad)_{<0}$ denotes the projection 
to the negative power part 
\[
  \left(\sum_{n\in\ZZ}a_n\Lambda^n\right)_{<0}
  = \sum_{n<0}a_n\Lambda^n. 
\]

The Lax equations can be converted to the evolution equations 
\[
  \frac{\rd W}{\rd t_k} 
  = (W\Lambda^k W^{-1})_{\ge 0}W - W\Lambda^k 
  = - (W\Lambda^k W^{-1})_{<0}W 
\]
for the dressing operator 
\[
  W = 1 + \sum_{n=1}^\infty w_n\Lambda^{-n}. 
\]
$L$ is thereby expressed in the dressed form 
\[
  L = W\Lambda W^{-1}.
\]
The wave function 
\[
  \Psi = \left(1 + \sum_{n=1}^\infty w_nz^{-n}\right)
         \exp\left(\sum_{k=1}^\infty t_kz^k\right)
\]
satisfies the auxiliary linear equations 
\[
  \frac{\rd\Psi}{\rd t_k} = B_k\Psi,\quad 
  L\Psi = z\Psi. 
\]
The amplitude part of $\Psi$ is related 
to the tau function $\calT(s,\bst)$ 
\footnote{For consistency with our previous work, 
the variable $s$ of the tau function is shifted to $s - 1$.} 
as 
\[
  1 + \sum_{n=1}^\infty w_nz^{-n} 
  = \frac{\calT(s-1,\bst - [z^{-1}])}{\calT(s-1,\bst)},\quad 
  [z] = \left(\frac{z^k}{k}\right)_{k=1}^\infty. 
\]

We can define the logarithm $\log L$ 
and the fractional power $L^\alpha$ of $L$ as 
\[
  \log L = W\log\Lambda W^{-1},\quad 
  L^\alpha = W\Lambda^\alpha W^{-1}.
\]
Since 
\[
  \log\Lambda = \rd_s,\quad 
  \Lambda^\alpha = e^{\alpha\rd_s}, 
\]
$\log L$ is a differential-difference operator 
\[
  \log L = \rd_s - \frac{\rd W}{\rd s}W^{-1}, 
\]
and $L^\alpha$ is a difference operator with fractional shift 
\[
  L^\alpha 
  = W\cdot W(s+\alpha)^{-1}e^{\alpha\rd_s} 
  = (1 + a_1\Lambda^{-1} + a_2\Lambda^{-2} + \cdots)e^{\alpha\rd_s}, 
\]
where $W(s+\alpha)$ means substituting $s \to s + \alpha$ 
in the coefficients of $W$.

\section{Hypergeometric tau functions}

\subsection{General form}

For the moment, let us restrict the value of $s$ to $\ZZ$. 
We call the tau function 
\beq
  \calT(s,\bst) 
   = \sum_{\lambda\in\calP}S_\lambda(\bst)h_\lambda(s)S_\lambda(\bsc) 
  \label{HGtau}
\eeq
a hypergeometric tau function of the lattice KP hierarchy. 
Let us specify the notations used here.  
\begin{itemize}
\item
$\calP$ denotes the set of all partitions 
$\lambda = (\lambda_i)_{i=1}^\infty$, 
$\lambda_1\ge\lambda_2\ge\cdots\ge\lambda_{l(\lambda)} > 0$, 
$\lambda_i = 0$ for $i > l(\lambda)$. 
\item 
$S_\lambda(\bst)$'s are the Schur polynomials 
\[
\begin{gathered}
  S_\lambda(\bst) = \det(S_{\lambda_i-i+j}(\bst))_{i,j=1}^\infty,\\
  \sum_{n=0}^\infty S_n(\bst)z^n = \exp\left(\sum_{k=1}^\infty t_kz^k\right). 
\end{gathered}
\]
The determinant is understood to be the stable limit 
\[
\begin{aligned}
  \det(S_{\lambda_i-i+j}(\bst))_{i,j=1}^\infty 
  &= \lim_{n\to\infty}\det(S_{\lambda_i-i+j}(\bst))_{i,j=1}^n\\
  &= \det(S_{\lambda_i-i+j}(\bst))_{i,j=1}^n \quad 
     \text{for $n \ge l(\lambda)$}  
\end{aligned}
\]
of the $n$-th principal minor.  By substituting 
\[
  t_k = \frac{1}{k}\sum_{i=1}^\infty x_i^k,
\]
$S_\lambda(\bst)$'s turn into the Schur functions 
$s_\lambda(\bsx)$, $\bsx = (x_k)_{k=1}^\infty$, 
in the theory of symmetric functions \cite{Mac-book}. 
\item 
$\bsc$ is a set of constants $\bsc = (c_k)_{k=1}^\infty$. 
\item 
$h_\lambda(s)$'s are defined by a diagonal matrix 
$\diag(h_n)_{n\in\ZZ}$ of non-zero constants as 
\beq
  h_\lambda(s) 
  = \frac{\prod_{i=1}^\infty h_{\lambda_i-i+s+1}}{\prod_{i=1}^\infty h_{-i+1}}.
  \label{h_lam(s)-def}
\eeq
The right hand side is a somewhat formal expression. 
Actually, since all but a finite number of terms 
in the numerator and the denominator pairwise cancel out, 
this formal expression can be reduced to a finite expression, 
e.g., 
\beq
  h_\emptyset(s) = \begin{cases}
    \prod_{n=1}^s h_n & \text{for $s > 0$},\\
    1 & \text{for $s = 0$},\\
    \prod_{n=s+1}^0 h_n^{-1} & \text{for $s < 0$}.
    \end{cases}
  \label{h_0(s)-def}
\eeq
Moreover, the ratio $h_\lambda(s)/h_\emptyset(s)$ 
can be cast into the so called contents product 
\beq
  \frac{h_\lambda(s)}{h_\emptyset(s)} 
      = \prod_{(i,j)\in\lambda}r_{j-i+s+1},\quad 
  r_n = \frac{h_n}{h_{n-1}}, 
  \label{h_lam(s)-contents}
\eeq
where $(i,j)\in\lambda$ means that $(i,j)$ is a box 
in the $i$-th row and the $j$-th column 
of the Young diagram of shape $\lambda$. 
\end{itemize}

(\ref{HGtau}) is a specialization of the tau function 
\beq
  \calT(s,\bst,\bar{\bst}) 
   = \sum_{\lambda\in\calP}S_\lambda(\bst)h_\lambda(s)S_\lambda(-\bar{\bst}) 
  \label{TodaHGtau}
\eeq
of the 2D Toda hierarchy to $\bar{\bst} = - \bsc$.  
It is the tau functions of this form that are studied 
by Orlov and Scherbin \cite{OS01a,OS01b}. 
The definition (\ref{h_lam(s)-def}) of $h_\lambda(s)$'s 
stems from a fermionic description of the tau functions. 
The diagonal matrix $\diag(h_n)_{n\in\ZZ}$ corresponds 
to the element 
\[
  h = \exp\left(\sum_{n\in\ZZ}\log h_n:\psi_{-n}\psi^*_n:\right)
\]
of the $\mathrm{GL}(\infty)$ group that acts on 
the Fock space of 2D complex free fermion fields 
\[
  \psi(z) = \sum_{n\in\ZZ}\psi_nz^{-n-1},\quad 
  \psi^*(z) = \sum_{n\in\ZZ}\psi^*_nz^{-n}. 
\]
$h_\lambda(s)$'s are the diagonal matrix elements 
$\langle \lambda,s|h|\lambda,s\rangle$ of $h$ 
with respect to a basis $|\lambda,s\rangle$, 
$\lambda\in\calP$, $s \in \ZZ$, of the Fock space. 
$|s\rangle = |\emptyset,s\rangle$ is the ground state 
in the charge-$s$ sector, and the excited states 
are labelled by the partition $\lambda$.  
Thus being a specialization of the tau function 
(\ref{TodaHGtau}) of the 2D Toda hierarchy, 
(\ref{HGtau}) becomes a tau function of the lattice KP hierarchy.

\subsection{Specialization of $h_n$'s}

Our consideration in this paper is focussed on the special case 
\beq
  h_n = \exp\left(\frac{\beta}{2}\left(n - \frac{1}{2}\right)^2
        \right)Q^{n-1/2}, 
  \label{special-hn}
\eeq
where $\beta$ and $Q$ are non-zero constants.  
One can use (\ref{h_0(s)-def}) and (\ref{h_lam(s)-contents}) 
to compute $h_\lambda(s)$ explicitly as 
\beq
  h_\lambda(s) = \exp\left(\frac{\beta}{2}
    \left(\kappa(\lambda) + 2s|\lambda| + \frac{4s^3-s}{12}\right)
    \right) Q^{|\lambda|+s^2/2}. 
  \label{special-h_lam(s)}
\eeq
$|\lambda|$ and $\kappa(\lambda)$ denote 
the parts sum and twice the contents sum: 
\[
  |\lambda| = \sum_{i=1}^\infty\lambda_i,\quad 
  \kappa(\lambda) = 2\sum_{(i,j)\in\lambda}(j - i). 
\]
$\kappa(\lambda)$ has the following alternative expression: 
\[
  \kappa(s) 
  = \sum_{i=1}^\infty\lambda_i(\lambda_i - 2i + 1) 
  = \sum_{i=1}^\infty\left(\left(\lambda_i-i+\frac{1}{2}\right)^2 
       - \left(-i+\frac{1}{2}\right)^2 \right). 
\]

In the fermionic interpretation, the diagonal matrix 
$\diag(h_n)_{n\in\ZZ}$ amounts to the operator 
\beq
  h = e^{\beta K/2}Q^{L_0}, 
  \label{special-h}
\eeq
where 
\footnote{We have modified the definition of $L_0$ 
in our previous work.}
\[
\begin{aligned}
  K &= \sum_{n\in\ZZ}\left(n-\frac{1}{2}\right)^2:\psi_{-n}\psi^*_n:,\\
  L_0 &= \sum_{n\in\ZZ}\left(n-\frac{1}{2}\right):\psi_{-n}\psi^*_n:. 
\end{aligned}
\]
One can derive (\ref{special-h_lam(s)}) from the formulae 
\[
\begin{aligned}
  \langle\lambda,s|K|\lambda,s\rangle 
  &= \kappa(\lambda) + 2s|\lambda| + \frac{4s^3-s}{12},\\
  \langle\lambda,s|L_0|\lambda,s\rangle 
  &= |\lambda| + \frac{s^2}{2}
\end{aligned}
\]
of the matrix elements of $K$ and $L_0$. 
The tau function $\calT(s,\bst,\bar{\bst})$ 
of the 2D Toda hierarchy defined by the same $h$ 
is nothing but Okounkov's generating function 
of the double Hurwitz numbers \cite{Okounkov00}. 

Although we have assumed that the variable $s$ takes values 
in $\ZZ$, the expression of $h_\lambda(s)$ 
in (\ref{special-h_lam(s)}) is meaningful for $s \in \RR$ 
(or even for $s \in \CC$) as well .  
Actually, one can prove that the shifted function 
$\calT(s+a,\bst)$, too, is a tau function 
of the 2D Toda hierarchy for any constant $a$. 
\footnote{The proof is parallel to the case of 
the tau function of the Hodge integrals \cite{Takasaki19}.}. 
This implies that $\calT(s,\bst)$ is a tau function 
in the sense explained in Section 2.

\section{Factorization problem and its implications}

\subsection{Factorization problem and initial value of $W$}

The dressing operator $W$ can be captured 
by the factorization problem \cite{Takasaki84}
\beq
  \exp\left(\sum_{k=1}^\infty t_k\Lambda^k\right)U
  \exp\left(\sum_{k=1}^\infty c_k\Lambda^{-k}\right)
  = W^{-1}\bar{W}, 
  \label{factor-prob}
\eeq
where $\bar{W}$ is a difference operator of the form 
\[
  \bar{W} = \sum_{n=0}^\infty\bar{w}_n\Lambda^n, 
\]
and $U$ is an invertible operator that does not depend 
on $\bst$.  The problem is to find the pair $W,\bar{W}$ 
from the given data $U$.  This factorization problem 
is a specialization of the factorization problem 
for the 2D Toda hierarchy.  

In the case where $h_n$'s are specialized as shown 
in (\ref{special-hn}), $U$ is the difference operator 
\beq
  U = e^{\beta(s-1/2)^2/2}Q^{s-1/2} 
  \label{special-U}
\eeq
obtained from (\ref{special-h}) by the correspondence 
\[
\begin{gathered}
  K \leftrightarrow \diag((n-1/2)^2)_{n\in\ZZ} 
    \leftrightarrow (s - 1/2)^2,\\
  L_0 \leftrightarrow \diag(n-1/2)_{n\in\ZZ} 
    \leftrightarrow (s - 1/2)
\end{gathered}
\]
among the fermion bilinear operators, 
the infinite matrices and the difference operators.  

When $\bst$ is further specialized to $\bst = \bszero$, 
the factorization problem (\ref{factor-prob}) 
can be solved immediately, because the left hand side 
is then factorized in an almost final form.  
This yields the expression
\beq
  W_{\bszero} = U\exp\left(- \sum_{k=1}^\infty c_k\Lambda^{-k}\right)U^{-1} 
  \label{W0}
\eeq
of the initial value 
\[
  W_{\bszero} = W|_{\bst=\bszero}
\]
of the dressing operator. 

\subsection{Initial values of $L$, $L^\alpha$ and $\log L$}

We can use (\ref{W0}) to compute the initial value
\[
  L_{\bszero} = L|_{\bst=\bszero}
\]
of the Lax operator $L$ and its logarithm and fractional power 
as follows: 

\begin{prp}
\begin{align}
  L_{\bszero} 
  &= \Lambda\exp\left(\sum_{k=1}^\infty c_kQ^k(1 - e^{-\beta k})
                e^{-\beta k(k+1)/2}e^{\beta ks}\Lambda^{-k}\right),
     \label{L0}\\
  L_{\bszero}^\alpha 
  &= \Lambda^\alpha\exp\left(\sum_{k=1}^\infty c_kQ^k(1 - e^{-\alpha\beta k})
                e^{-\beta k(k+1)/2}e^{\beta ks}\Lambda^{-k}\right),
     \label{L0^alpha}\\
  \log L_{\bszero} &= \log\Lambda 
     + \beta\sum_{k=1}^\infty kc_kQ^ke^{-\beta k(k+1)/2}e^{\beta ks}\Lambda^{-k}.
     \label{logL0}
\end{align}
\end{prp}

\proof
Let us rewrite (\ref{W0}) using (\ref{special-U}) as 
\[
  W_{\bszero} = e^{\beta(s-1/2)^2/2}
    \exp\left(- \sum_{k=1}^\infty c_kQ^k\Lambda^{-k}\right)e^{-\beta(s-1/2)^2/2}
\]
and compute 
\[
  L_{\bszero} = W_{\bszero}\Lambda W_{\bszero}^{-1} 
\]
step by step.  Since 
\[
  e^{-\beta(s-1/2)^2/2}\Lambda e^{\beta(s-1/2)^2/2}
  = e^{-\beta(s-1/2)^2/2}e^{\beta(s+1/2)^2/2}\Lambda 
  = e^{\beta s}\Lambda 
\]
and 
\[
\begin{aligned}
  &\exp\left(- \sum_{k=1}^\infty c_kQ^k\Lambda^{-k}\right) 
  e^{\beta s}\Lambda\exp\left(\sum_{k=1}^\infty c_kQ^k\Lambda^{-k}\right)\\
  &= e^{\beta s}\Lambda
    \exp\left(\sum_{k=1}^\infty c_kQ^k(1 - e^{-\beta k})\Lambda^{-k}\right), 
\end{aligned}
\]
$L_{\bszero}$ can be expressed as 
\[
  L_0 = e^{\beta(s-1/2)^2/2}e^{\beta s}\Lambda 
    \exp\left(\sum_{k=1}^\infty c_kQ^k(1 - e^{-\beta k})\Lambda^{-k}\right)\\
    e^{-\beta(s-1/2)^2/2}. 
\]
Moreover, since 
\[
  e^{\beta(s-1/2)^2/2}e^{\beta s}\Lambda e^{-\beta(s-1/2)^2/2} = \Lambda 
\]
and 
\beq
  e^{\beta(s-1/2)^2/2}\Lambda^{-k}e^{-\beta(s-1/2)^2/2} 
  = e^{-\beta k(k+1)/2}e^{\beta ks}\Lambda^{-k}, 
  \label{beta-Lam-rel}
\eeq
the last expression of $L_{\bszero}$ boils down to (\ref{L0}). 
The fractional power 
\[
  L_{\bszero}^\alpha = W_{\bszero}\Lambda^\alpha W_{\bszero}^{-1}
\]
can be treated in much the same way and cast into 
the expression shown in (\ref{L0^alpha}).  Lastly, 
we can compute $\log L_{\bszero}$ from (\ref{L0^alpha}) as 
\[
   \log L_{\bszero} 
  = \left.\frac{\rd}{\rd\alpha}L_{\bszero}^\alpha\right|_{\alpha=0}. 
\]
This yields (\ref{logL0}). 
\qed

\begin{rem}
One can use (\ref{beta-Lam-rel}) to rewrite (\ref{logL0}) as 
\begin{align}
  L_{\bszero}^\alpha &= \Lambda^\alpha e^{\beta(s-1/2)^2/2} 
     \exp\left(\sum_{k=1}^\infty c_kQ^k(1 - e^{-\alpha\beta k})
     \Lambda^{-k}\right)e^{-\beta(s-1/2)^2/2},
  \label{L0^alpha-bis}\\
  \log L_{\bszero} &= \log\Lambda 
    + \beta e^{\beta(s-1/2)^2/2}\sum_{k=1}^\infty kc_kQ^k\Lambda^{-k}
      e^{-\beta(s-1/2)^2/2}. 
  \label{logL0-bis}
\end{align}
\end{rem}

\subsection{Special values of $\bsc$}

Inspired by the work of Orlov \cite{Orlov05}, 
we choose the following four values 
of $\bsc = (c_k)_{k=1}^\infty$ as the base points 
of our quest for new integrable structures: 
\begin{itemize}
\item[(a)]
$\bsc = \bst_\infty = (1,0,0,\ldots)$. 
\item[(b)]
$\bsc = \bst(a) = \left(\dfrac{a}{k}\right)_{k=1}^\infty$.
\item[(c)]
$\bsc = \bst(\infty,q) = \left(\dfrac{1}{k(1-q^k)}\right)_{k=1}^\infty$. 
\item[(d)]
$\bsc = \bst(a,q) = \left(\dfrac{1-q^{ak}}{k(1-q^k)}\right)_{k=1}^\infty$. 
\end{itemize}
$a$ and $q$ are non-zero constants, 
and $q$ is assumed to be in the range $|q| < 1$.  
The Schur polynomials are known to take particular values 
at these special points \cite{Mac-book}.  
For example, 
\[
  S_\lambda(\bst_\infty) 
  = \frac{\dim\lambda}{|\lambda|!} 
  = \prod_{(i,j)\in\lambda}\frac{1}{h(i,j)}. 
\]
where $\dim\lambda$ is the dimension 
of the irreducible representation 
of the symmetric group $S_d$, $d = |\lambda|$, 
of type $\lambda$, in other words, 
the number of standard Young tableaux of shape $\lambda$. 
$h(i,j)$ is the hook length of the box $(i,j)$ 
of the Young diagram.  $S_\lambda(\bst(\infty,q))$ 
is a $q$-deformation of $S_\lambda(\bst_\infty)$: 
\[
  S_\lambda(\bst(\infty,q)) 
  = \frac{q^{-\kappa(\lambda)/4-|\lambda|/2}}
    {\prod_{(i,j)\in\lambda}(q^{-h(i,j)/2} - q^{h(i,j)/2})}. 
\]
Moreover, $S_\lambda(\bst(a))$ and $S_\lambda(\bst(a,q)$ 
are obtained from these values by multiplying simple factors 
(which are also contents product). 

In each case, the initial value $L_{\bszero}$ 
of the Lax operator turns out to take a significant form, 
which indicates what kind of integrable structure 
is hidden behind.  

\begin{itemize}
\item[(a)]
This case is related to the single Hurwitz numbers 
of $\CC\PP^1$ \cite{Okounkov00,HO15}.  
$\log L_{\bszero}$ can be computed 
with the aid of (\ref{logL0}) as 
\beq
  \log L_{\bszero} = \log\Lambda + \beta Qe^{\beta(s-1)}\Lambda^{-1}.
  \label{logL0(a)}
\eeq
As shown in our previous work \cite{Takasaki18b}, 
this differential-difference operator is related 
to the continuum limit \cite{Bog88,Itoh88} 
of the Bogoyavlensky-Itoh-Narita lattice 
\cite{Narita82,Bog87,Itoh87}. 
\item[(b)]
This case is a kind of one-parameter deformation of case (a). 
$\log L_{\bszero}$ can be computed with the aid of 
(\ref{logL0-bis}) as 
\beq
  \log L_{\bszero} = \log\Lambda 
    + \beta aQe^{\beta(s-1)}\Lambda^{-1}(1 - Qe^{\beta(s-1)}\Lambda^{-1})^{-1} 
  \label{logL0(b)}
\eeq
(\ref{logL0(a)}) can be recovered from this operator 
in the scaling limit as $a \to \infty$, $Q \to 0$ 
and $aQ$ is fixed to a non-zero constant (which becomes 
the new constant $Q$). 
\item[(c)]
This case is related to the cubic Hodge integrals 
and the topological vertex \cite{GV99,MV01,AKMV03} 
if $Q$ and $\beta$ are chosen as 
\beq
  Q = q^{1/2},\quad \beta = (f+1)\log q,
  \label{special-Qbeta}
\eeq
where $f$ is a non-negative integer called the framing number.  
After some algebra based on (\ref{L0^alpha-bis}), 
the fractional power of $L_{\bszero}$ of order $1/(f+1)$ 
turns out to take the simple form 
\beq
  L_{\bszero}^{1/(f+1)} 
  = (1 - q^{(f+1)(s-1)+1/2}\Lambda^{-1})\Lambda^{1/(f+1)}. 
  \label{L0^alpha(c)}
\eeq
As pointed out in our recent work \cite{Takasaki19}, 
this difference operator of fractional order 
(for $f > 0$ 
\footnote{The case where $f = 0$ is trivial 
in the context of integrable structures.})
is related to the Bogoyavlensky-Itoh-Narita lattice. 
The case of $f = 1$ amounts to the Volterra lattice.  
This interpretation can be extended to the case 
where $f$ is a positive rational number. 
Actually, in the context of the Hodge integrals 
\cite{LLZ0306,OP0307}, $f$ can be replaced 
by an arbitrary real value $\tau \not= 0,-1$. 
The counterpart 
\beq
  L_{\bszero}^{1/(\tau+1)} 
  = (1 - q^{(\tau+1)(s-1)+1/2}\Lambda^{-1})\Lambda^{1/(\tau+1)}. 
  \label{L0^alpha(c)bis}
\eeq
of the foregoing expression (\ref{L0^alpha(c)}) 
persists to hold as far as $\tau > -1$, in particular, 
in the new regime $-1 < \tau < 0$. 
If $\tau$ is a rational number in this negative interval, 
one can choose two positive coprime integers $M,N$ with $M < N$ 
such that 
\[
  \frac{1}{\tau+1} = \frac{N}{M}. 
\]
(\ref{L0^alpha(c)bis}) then implies that the $N$-th power 
of $L_{\bszero}$ takes the special form 
\beq
  L_{\bszero}^N = \Lambda^N + v_{1\bszero}\Lambda^{N-1} 
     + \cdots + v_{M\bszero}
\Lambda^{N-M}.
  \label{L0^N(c)}
\eeq
The regime $-1 < \tau < 0$ is thus related 
to the lattice version 
of the Gelfand-Dickey hierarchy \cite{Frenkel95}. 

\item[(d)]
This case, as well as case (c), is related to 
topological string theory on toric Calabi-Yau threefolds 
\cite{TN15,TN16,Takasaki13b}.  The target space therein 
is the resolved conifold, whereas that of case (c) is $\CC^3$.  
The extra factor $q^a$ in the definition of $\bst(a,q)$ 
is a geometric parameter of the resolved conifold.  
Under the same specialization (\ref{special-Qbeta}) as case (c), 
the fractional power of $L_{\bszero}$ of order $1/(f+1)$ 
develops yet another operator factor: 
\beq
  L_{\bszero}^{1/(f+1)} 
  = (1 - q^{(f+1)(s-1)+1/2}\Lambda^{-1})
    (1 - q^{(f+1)(s-1)+a+1/2}\Lambda^{-1})^{-1}\Lambda^{1/(f+1)}. 
  \label{L0^alpha(d)}
\eeq
(\ref{L0^alpha(c)}) can be recovered from this operator 
in the limit as $a \to \infty$.  In the case where $f = 0$, 
(\ref{L0^alpha(d)}) emerges in the Ablowitz-Ladik 
(or relativistic Toda) hierarchy \cite{BCR11,Takasaki13a}. 
\end{itemize}

Thus the expressions of $L_{\bszero}$ or $\log L_{\bszero}$ 
indicate the existence of various underlying integrable structures.  
These integrable structures should be realized as reductions 
of the lattice KP hierarchy.  The reduction conditions 
on the Lax operator have to be consistent with the flows 
of the lattice KP hierarchy.  

In the subsequent sections, we present these reductions 
in a general form and check the consistency.

\section{Reductions of Gelfand-Dickey type}

\subsection{Lattice version of Gelfand-Dickey hierarchy}

The reductions of this type are defined by the condition 
that a positive integral power of the Lax operator 
consists of a finite number of terms: 
\beq
  L^N = \Lambda^N + v_1\Lambda^{N-1} + \cdots + v_M\Lambda^{N-M},\quad 
  v_n = v_n(s,\bst). 
  \label{L^N-GD}
\eeq
Following Frenkel's work \cite{Frenkel95}
\footnote{Frenkel uses $q$-shift operators rather than 
shift operators.  This is not an essential difference.}, 
we call this reduction a lattice version 
of the Gelfand-Dickey hierarchy. 

If $M$ is greater than $N$, this reduction is a subsystem 
of the bigraded Toda hierarchy \cite{Carlet06} 
obtained by suppressing the flows with respect 
to $\bar{\bst} = (\bar{t}_1,\bar{t}_2,\ldots)$. 
On the other hand, as mentioned in Sect. 4.3, 
a Lax operator of this type with $M < N$ emerges 
in the context of the cubic Hodge integrals.  

One can directly verify  that (\ref{L^N-GD}) 
is a consistent reduction condition, namely, preserved 
by the flows of the lattice KP hierarchy, for any $M > 0$.  
The reasoning is based on the Lax equations 
\beq
  \frac{\rd\calL}{\rd t_k} = [B_k,\calL] = [\calL,B_k^{-}]
  \label{red-Laxeq}
\eeq
satisfied by $\calL = L^N$.  The two expressions 
of the Lax equations imply that the right hand side 
is an operator of the form 
\[
  [B_k,\calL] = [\calL,B_k^{-}] 
  = f_{1k}\Lambda^{N-1} + \cdots + f_{Mk}\Lambda^{N-M}. 
\]
The Lax equations thus turn out to be equivalent 
to the evolution equations 
\[
  \frac{\rd v_n}{\rd t_k} = f_{nk},\quad n = 1,\ldots,M, 
\]
for the coefficients $v_1,\ldots,v_M$ of $L^N$
\footnote{To be more precise, we encounter the problem 
of non-locality in the definition of $B_k$'s.  A remedy is 
to introduce a parameter (Planck constant) $\epsilon$ 
in the construction of the hierarchy \cite{Carlet06}.}.
Therefore, if the initial value of $L^N$ at an initial time 
(e.g., $\bst = \bszero$) is an operator of the form 
(\ref{L^N-GD}), $L^N$ itself takes the same form at all times.

It may happen, however, that $v_1,\ldots,v_M$ are not 
the full degrees of freedom of the reduced system 
\footnote{This is also related to the non-locality of the flows.}.  
An extreme situation arises in the case where $M = 0$ 
--- no degree of freedom is left in $L^N = \Lambda^N$. 
Of course, all degrees of freedom are carried by $L$ itself.  
In particular, as we show below, the coefficient $u_1$ of $L$ 
is related to the usual Gelfand-Dickey hierarchy 
in a continuous space.

\subsection{Relation to usual Gelfand-Dickey hierarchy}

Let us consider the spectral problem 
\beq
  (\Lambda^N + v_1\Lambda^{N-1} + \cdots + v_M\Lambda^{N-M})\Psi 
  = L^N\Psi = z^N\Psi
  \label{lGD-spec}
\eeq
implied by the reduction condition (\ref{L^N-GD}).  
One can use the auxiliary linear equation 
\[
  \frac{\rd\Psi}{\rd t_1} = B_1\Psi = (\Lambda + u_1)\Psi 
\]
of the first flow to rewrite the terms $\Lambda^m\Psi$'s 
on the left hand side of (\ref{lGD-spec}) as 
\[
  \Lambda^m\Psi = (\rd_1 - u_1(s+m-1))\cdots (\rd_1 - u_1(s))\Psi,\quad 
  \rd_1 = \rd/\rd t_1. 
\]
Thus (\ref{lGD-spec}) turn into the spectral problem 
\[
  (\rd_1^N + a_2\rd_1^{N-2} + \cdots + a_N)\Psi 
  = z^N\Psi  
\]
of the Gelfand-Dickey type.  The auxiliary linear equations 
of the other flows can be similarly transformed 
to differential equations of the form 
\[
  \frac{\rd\Psi}{\rd t_k} = \tilde{B}_k\Psi, \quad 
  \tilde{B}_k = \rd_1^k + \tilde{b}_{k2}\rd_1^{k-2} 
     + \cdots + \tilde{b}_{kk}. 
\]
These equations can be identified with the auxiliary linear equations 
of the usual Gelfand-Dickey hierarchy, in which $t_1$ plays 
the role of the spatial variable.

\section{Reductions of Volterra type}

\subsection{Bogoyavlensky-Itoh-Narita lattice and its continuum limit}

Let us formulate the reduction to 
the Bogoyavlensky-Itoh-Narita lattice 
\cite{Narita82,Bog87,Itoh87} 
in a form generalizing (\ref{L0^alpha(c)}).  
The reduction can be characterized by the condition 
that the fractional power of $L$ of order $1/(f+1)$ 
consists of two terms: 
\beq
  L^{1/(f+1)} = (1 - v\Lambda^{-1})\Lambda^{1/(f+1)} 
  = \Lambda^{1/{f+1}} - v\Lambda^{-f/(f+1)},\quad 
  v = v(s,\bst). 
  \label{L^alpha-BIN}
\eeq
The Lax equations for $\calL = L^{1/(f+1)}$ 
take the same form as (\ref{red-Laxeq}), 
and implies that $v$ satisfies the evolution equations 
\[
  \frac{\rd v}{\rd t_k} = f_k, 
\]
where $f_k$'s are the coefficients of 
\[
  [B_k,\calL] = [\calL,B_k^{-}] = f_k\Lambda^{-f/(f+1)}.
\]

The Bogoyavlensky-Itoh-Narita lattice is a ``hungry'' 
\footnote{This term was coined by S.~Tsujimoto.}
generalization of the well known Volterra lattice. 
Bogoyavlensky and Itoh also considered 
its continuum limit as $f \to \infty$ \cite{Bog88,Itoh88}.  
The continuum version can be characterized 
in the lattice KP hierarchy by the reduction condition 
\beq
  \log L = \log\Lambda - \tilde{v}\Lambda^{-1},\quad 
  \tilde{v} = \tilde{v}(s,\bst). 
  \label{logL-BI}
\eeq
The continuum limit is actually a scaling limit 
letting $f \to \infty$ and keeping $\tilde{v} = (f+1)v$ finite. 
The terms $\log L$ and $\log\Lambda$ in (\ref{logL-BI}) 
emerge in the scaling limit as 
\[
  \log L = \lim_{f\to\infty}(f+1)(L^{1/(f+1)} - 1),\quad 
  \log \Lambda = \lim_{f\to\infty}(f+1)(\Lambda^{1/(f+1)} - 1). 
\]
We have seen in Section 4.3 that the hypergeometric 
tau function specialized to $\bsc = \bst_\infty$ 
yields the operator (\ref{logL0(a)}) of this form. 

Note that $L$, hence $B_k$'s, in these two reductions 
can be expressed in terms of $v$ and its $s$-shifts only. 
The right hand side of the evolution equations for $v$ 
are local in this sense.  Thus $v$ carries full degrees 
of freedom of the reduced system. This is in contrast 
with the reduction of the Gelfand-Dickey type. 

\subsection{Generalizations}

We find further generalizations of these reductions 
of the Volterra type.  Let us consider the generalization 
\[
  \bsc = (c_1,\ldots,c_N,0,0,\ldots)
\]
of the special value $\bsc = \bst_\infty$ of Sect. 4.3. 
The associated hypergeometric tau functions 
(in particular, with $c_1 = \cdots = c_{N-1} = $ and $c_N =1$)
are known to be related to generalized Hurwitz numbers 
\cite{JPT08,MSS13}. 
(\ref{logL0}) then implies that $\log L_{\bszero}$ 
is an operator of the form 
\[
  \log L_{\bszero} = \log\Lambda + \sum_{k=1}^N v_{k\bszero}\Lambda^{-k}. 
\]
This indicates the existence of a reduction defined 
by the condition 
\beq
  \log L = \log\Lambda + \sum_{n=1}^N v_n\Lambda^{-n},\quad 
  v_n = v_n(s,\bst). 
  \label{logL-gBI}
\eeq
The consistency of this reduction condition can be verified 
in the same manner as the case of (\ref{logL-BI}). 

In much the same way, the special value $\bsc = \bst(\infty,q)$ 
of Section 4.3 can be generalized as 
\[
  c_k = \frac{1}{k(1-q^k)}\sum_{n=1}^N q^{b_nk}, 
\]
where $b_n$'s are new constants.  If $Q$ and $\beta$ 
are chosen as shown in (\ref{special-Qbeta}), 
the associated fractional power of $L_{\bszero}$ 
of order $1/(f+1)$ turns out to take the product form 
\[
  L_{\bszero}^{1/(f+1)}
  = \prod_{n=1}^N(1 - q^{(f+1)(s-1)+b_n+1/2}\Lambda^{-1})
    \cdot\Lambda^{1/(f+1)}. 
\]
We are thus led to the possibility of a new type 
of reduction defined by the condition 
\beq
  L^{1/(f+1)} = B\Lambda^{1/(f+1)},
  \label{L^alpha-gBIN}
\eeq
where $B$ is a difference operator of the form 
\[
  B = 1 + \sum_{n=1}^N v_n\Lambda^{-n},\quad 
  v_n = v_n(s,\bst). 
\]
The foregoing reduction condition (\ref{logL-gBI}) 
can be derived from (\ref{L^alpha-gBIN}) 
by a suitable scaling limit as $f \to \infty$.  
The consistency of the reduction condition 
(\ref{L^alpha-gBIN}) will be explained 
in a broader context below.

\section{Variation of rational reductions}

\subsection{Reduction conditions}

The initial value (\ref{L0^alpha(d)}) 
for the special value $\bsc = \bst(a,q)$ in Section 4.3 
suggests a variation of (\ref{L^alpha-BIN}) of the form 
\beq
  L^{1/(f+1)} 
  = (1 - v\Lambda^{-1})(1 - u\Lambda^{-1})^{-1}\Lambda^{1/(f+1)} 
  \label{L^alpha-AL}
\eeq
with one more field $u = u(s,\bst)$.  
This also is a variation of the reduction condition 
for the Ablowitz-Ladik hierarchy \cite{BCR11,Takasaki13a}.  

The Ablowitz-Ladik reduction is a special case 
of the so called ``rational reductions'' 
of the 2D Toda hierarchy \cite{BCRR14}.  
The consistency of the reduction condition 
(\ref{L^alpha-AL}) can be verified by the same method 
as used for those rational reductions.  
Moreover, from the point of view of rational reductions, 
(\ref{L^alpha-AL}) can be further generalized as follows.  

Let us recall that a generalization 
of the Ablowitz-Ladik hierarchy emerges 
in topological string theory on a generalization 
of the resolved conifold \cite{Takasaki13b,BCRR14}. 
This case amounts to choosing the constants $\bsc$ as 
\[
  c_k = \frac{1}{k(1-q^k)}
       \left(\sum_{n=1}^N q^{b_nk} - \sum_{n=1}^N q^{a_nk}\right). 
\]
$q^{b_n}$ and $q^{a_n}$ are geometric parameters 
of the target space geometry.  The associated 
fractional power $L_{\bszero}^{1/(f+1)}$ of $L_{\bszero}$ 
then takes the ``rational'' form 
\[
\begin{aligned}
  L_{\bszero}^{1/(f+1)}
  &= \prod_{i=1}^N(1 - q^{(f+1)(s-1)+b_n+1/2}\Lambda^{-1})\\
  &\quad\mbox{}
     \cdot\prod_{n=1}^N(1 - q^{(f+1)(s-1)+a_n+1/2}\Lambda^{-1})^{-1}
     \cdot\Lambda^{1/(f+1)}. 
\end{aligned}
\]

This example suggest to consider a generalization 
of (\ref{L^alpha-AL}) of the form 
\beq
  L^{1/(f+1)} = BC^{-1}\Lambda^{1/(f+1)},
  \label{L^alpha-RR}
\eeq
where $B$ is the same difference operator as used 
in (\ref{L^alpha-gBIN}), and $C$ is another operator 
\footnote{Note that the coefficients $u_n$ are different 
from those of $L$.}
\[
  C = 1 + \sum_{n=1}^N u_n\Lambda^{-n},\quad 
  u_n = u_n(s,\bst). 
\]

\subsection{Consistency of reduction condition}

To verify the consistency of the reduction condition 
(\ref{L^alpha-RR}), we prove a few propositions. 

Let us introduce the difference operators 
\[
\begin{gathered}
  P_k = \left((BC^{-1}\Lambda^{1/(f+1)})^{k(f+1)}\right)_{\ge 0} = (L^k)_{\ge 0},\\
  Q_k = \left((C^{-1}\Lambda^{1/(f+1)}B)^{k(f+1)}\right)_{\ge 0} = (B^{-1}L^kB)_{\ge 0},\\
  R_k = \left((\Lambda^{1/(f+1)}BC^{-1})^{k(f+1)}\right)_{\ge 0} 
      = (\Lambda^{1/(f+1)}L^k\Lambda^{-1/(f+1)})_{\ge 0}
\end{gathered}
\]
for $k = 1,2,\ldots$.  

\begin{prp}
$P_k$ and $R_k$ satisfy the identity 
\beq
  R_k\Lambda^{1/(f+1)} = \Lambda^{1/(f+1)}P_k. 
  \label{RPLam-rel}
\eeq
\end{prp}

\proof
We have the general operator identity 
\[
\begin{aligned}
  \quad\left(\Lambda^{1/(f+1)}\cdot\sum_{n\in\ZZ}a_n(s)\Lambda^n
   \cdot\Lambda^{-1/(f+1)}\right)_{\ge 0}
  &= \sum_{n\ge 0}a_n(s+1/(f+1))\Lambda^n\\
  &= \Lambda^{1/(f+1)}\cdot\sum_{n\ge 0}a_n(s)\Lambda^n
     \cdot\Lambda^{-1/(f+1)}\\
  &=\Lambda^{1/(f+1)}\left(\sum_{n\in\ZZ}a_n(s)\Lambda^n
    \right)_{\ge 0}\cdot\Lambda^{-1/(f+1)}. 
\end{aligned}
\]
Consequently, 
\[
  R_k = \Lambda^{1/(f+1)}(L^k)_{\ge 0}\Lambda^{-1/(f+1)} 
      = \Lambda^{1/(f+1)}P_k\Lambda^{-1/(f+1)}. 
\]
\qed

\begin{prp}
If $B$ and $C$ satisfy the equations 
\beq
  \frac{\rd B}{\rd t_k} = P_kB - BQ_k,\quad 
  \frac{\rd C}{\rd t_k} = R_kC - CQ_k, 
  \label{BC-Laxeq}
\eeq
$L$ satisfies the Lax equations 
of the lattice KP hierarchy.  
\end{prp}

\proof
Differentiating $\calL = L^{1/(f+1)}$ with respect 
to $t_k$ yields
\[
  \frac{\rd\calL}{\rd t_k} 
  = \frac{\rd B}{\rd t_k}C^{-1}\Lambda^{1/(f+1)} 
    - BC^{-1}\frac{\rd C}{\rd t_k}C^{-1}\Lambda^{1/(f+1)}. 
\]
By (\ref{BC-Laxeq}) and (\ref{RPLam-rel}), 
one can rewrite the right hand side as 
\[
  \frac{\rd\calL}{\rd t_k} 
  = P_k\calL - \calL\Lambda^{-1/(f+1)}R_k\Lambda^{1/(f+1)} 
  = [P_k,\calL]. 
\]
Since $P_k = (L^k)_{\ge 0}$, these equations imply 
that $L = \calL^{f+1}$ satisfies the Lax equations 
of the lattice KP hierarchy. 
\qed

The Lax equations of the lattice KP hierarchy 
can be thus reduced to the equations (\ref{BC-Laxeq}) 
for $B$ and $C$.  Moreover, we can rewrite 
these equations as 
\[
  \frac{\rd B}{\rd t_k} = BQ_k^{-} - P_k^{-}B,\quad 
  \frac{\rd C}{\rd t_k} = CQ_k^{-} - R_k^{-}C, 
\]
where 
\[
\begin{gathered}
  P_k^{-} = \left((BC^{-1}\Lambda^{1/(f+1)})^{k(f+1)}\right)_{<0},\\
  Q_k^{-} = \left((C^{-1}\Lambda^{1/(f+1)}B)^{k(f+1)}\right)_{<0},\\
  R_k^{-} = \left((\Lambda^{1/(f+1)}BC^{-1})^{k(f+1)}\right)_{<0}
\end{gathered}
\]
These equations and (\ref{BC-Laxeq}) imply that 
the right hand side of (\ref{BC-Laxeq}) are 
linear combinations of $\Lambda^{-1},\ldots,\Lambda^{-N}$. 
Thus (\ref{BC-Laxeq}) can be reduced to evolution equations 
of the form 
\[
  \frac{\rd v_n}{\rd t_k} = f_{nk},\quad 
  \frac{\rd u_n}{\rd t_k} = g_{nk} 
\]
for the coefficients of $B$ and $C$.  
The right hand side of these equations are local 
because the coefficients of $L$, hence those 
of $P_k$, $Q_k$ and $R_k$, can be expressed 
in terms of $v_n$, $u_n$ and their $s$-shifts.  
This completes the verification of the consistency 
of the reduction condition (\ref{L^alpha-RR}). 

As a corollary, we find the consistency of 
the reduction condition (\ref{L^alpha-gBIN}) as well.  
This condition amounts to the case where $C = 1$.  
The condition $C = 1$ is preserved by 
the time evolutions of $B$ and $C$ under (\ref{BC-Laxeq}). 
The reduced equations for $B$ thereby take the form 
\[
  \frac{\rd B}{\rd t_k} = P_kB - BQ_k,
\]
with apparently the same definition of $P_k$ and $Q_k$ 
as those of (\ref{BC-Laxeq}).  These equations 
can be reduced to evolution equations of $v_n$'s.

\section{Further variation of rational reductions}

\subsection{Reduction condition}

We now turn to the problem of elucidating the status 
of (\ref{logL0(b)}).  This is the most mysterious case 
among the four specializations examined in Sect. 4.3. 
(\ref{logL0(b)}) looks like the initial value 
of an operator of the form 
\beq
  \log L = \log\Lambda - \tilde{u}\Lambda^{-1}(1 - u\Lambda^{-1})^{-1}.
  \label{logL-varAL}
\eeq
This may be thought of as a scaling limit 
of (\ref{L^alpha-AL}) as $f \to \infty$.  
Let us rewrite (\ref{L^alpha-AL}) as 
\[
    L^{1/(f+1)} 
  = \Lambda^{1/(f+1)} 
    - (v - u)\Lambda^{-1}(1 - u\Lambda^{-1})^{-1}\Lambda^{1/(f+1)}.
\]
Letting $f \to \infty$ while keeping $\tilde{u} = (f+1)(v - u)$ 
finite, we obtain (\ref{logL-varAL}) in the limit. 

The same idea can be applied to the more general 
reduction condition (\ref{L^alpha-RR}).  
We are thus led to the following reduction condition: 
\beq
  \log L = \log\Lambda + \tilde{C}C^{-1},
  \label{logL-varRR}
\eeq
where $\tilde{C}$ is an operator of the form 
\[
  \tilde{C} = \sum_{n=1}^N\tilde{u}_n\Lambda^{-n},\quad 
  \tilde{u}_n = \tilde{u}_n(s,\bst), 
\]
that amounts to the limit of $(f+1)(B - C)$ as $f \to \infty$.  

\subsection{Consistency of reduction condition}

The consistency of the reduction condition (\ref{logL-varRR}) 
can be verified with the aid of evolution equations 
for by $C$ and $\tilde{C}$.  Heuristically, those equations 
can be derived as a scaling limit of (\ref{BC-Laxeq}). 
The $t_k$-derivative of $\tilde{C}$ can be computed as 
\[
\begin{aligned}
  \frac{\rd\tilde{C}}{\rd t_k} 
  &= \lim_{f\to\infty}(f+1)\left(\frac{\rd B}{\rd t_k} 
       - \frac{\rd C}{\rd t_k}\right)\\
  &= \lim_{f\to\infty}(f+1)\left((P_k - R_k)B + R_k(B - C) 
       - (B - C)Q_k\right). 
\end{aligned}
\]
After some algebra, $C$ and $\tilde{C}$ turn out 
to satisfy the evolution equations 
\beq
  \frac{\rd\tilde{C}}{\rd t_k} = - \frac{\rd P_k}{\rd s}C 
      + P_k\tilde{C} - \tilde{C}Q_k,\quad 
  \frac{\rd C}{\rd t_k} = P_kC - CQ_k,
  \label{CC-Laxeq}
\eeq
where
\[
  P_k = (L^k)_{\ge 0}, \quad 
  Q_k = (C^{-1}L^kC)_{\ge 0}. 
\]

We can directly confirm the correctness of these equations. 

\begin{prp}
If $C$ and $\tilde{C}$ satisfy (\ref{CC-Laxeq}), 
$L$ satisfies the Lax equations of the lattice KP hierarchy. 
\end{prp}

\proof
The $t_k$-derivative of $\calL = \log\Lambda + \tilde{C}C^{-1}$ 
yields 
\[
  \frac{\rd\calL}{\rd t_k} 
  = \frac{\rd\tilde{C}}{\rd t_k}C^{-1} 
    - \tilde{C}C^{-1}\frac{\rd C}{\rd t_k}C^{-1}. 
\]
By (\ref{CC-Laxeq}), 
one can rewrite the right hand side as 
\[
  \frac{\rd\calL}{\rd t_k} 
  = - \frac{\rd P_k}{\rd s} + [P_k,\tilde{C}C^{-1}]
  = [P_k,\calL]. 
\]
Since $P_k = (L^k)_{\ge 0}$, these equations imply 
that $L = e^{\calL}$ satisfies the Lax equations 
of the lattice KP hierarchy. 
\qed

(\ref{CC-Laxeq}) can be cast into the dual form 
\beq
  \frac{\rd\tilde{C}}{\rd t_k} = \frac{\rd P_k^{-}}{\rd s}C 
      + \tilde{C}Q_k^{-} - P_k^{-}\tilde{C},\quad 
  \frac{\rd C}{\rd t_k} = CQ_k^{-} - P_k^{-}C, 
  \label{CC-Laxeq2}
\eeq
where 
\[
  P_k^{-} = (L^k)_{<0}, \quad 
  Q_k^{-} = (C^{-1}L^kC)_{<0}. 
\]
This equation is equivalent to (\ref{CC-Laxeq}) 
because of the commuation relation 
\[
  \frac{\partial L^k}{\partial s} 
  = [\log\Lambda,L^k] 
  = - [\tilde{C}C^{-1},L^k] 
\]
and the obvious identities 
\[
  P_k^{+} + P_k^{-} = L^k,\quad 
  Q_k^{+} + Q_k^{-} = C^{-1}L^kC.
\]
The commutation relation is a consequence of the relation 
\[
  \log L = \mathcal{L} = \log\Lambda + \tilde{C}C^{-1}. 
\]
(\ref{CC-Laxeq}) and (\ref{CC-Laxeq2}) show that 
(\ref{CC-Laxeq}) can be reduced to evolution equations 
for the coefficients $v_n,u_n$ of $C$ and $\tilde{C}$. 
The consistency of the reduction condition (\ref{logL-varRR}) 
can be thus verified.

\subsection*{Acknowledgements}

This work is a byproduct of collaboration with Toshio Nakatsu 
on topological vertex and Hodge integrals.  This work is also 
supported by the Research Institute for Mathematical Sciences, 
an International Joint Usage/Research Center located 
in Kyoto University.

\end{document}